\documentclass[
showpacs,preprintnumbers,amsmath,amssymb]{revtex4-2}
\usepackage{amsmath}
\usepackage{float}
\usepackage{graphicx}
\usepackage[]{caption}
\usepackage{tensor}
\usepackage[toc,page]{appendix}
\usepackage{color,soul}

\begin{document}

\title{PROPOSAL TO REPEAT THE ABRAHAM FORCE EXPERIMENT USING GIANT PERMITTIVITY MATERIALS }

\author {Iver Brevik\footnote{E-mail:iver.h.brevik@ntnu.no}}

\medskip

\affiliation{Department of Energy and Process Engineering, Norwegian University
of Science and Technology, N-7491 Trondheim, Norway}

 \today

\begin{abstract}

In the more than 100 years-old Abraham-Minkowski problem in macroscopic electrodynamics, the issue of how to observe the so-called Abraham term
${\bf f}^{\rm Aterm} = [(\varepsilon\mu-1)/c^2]
\partial/\partial t ({\bf E}\times {\bf H})$ has been a main point. Such a measurement is rather delicate. Recent years have seen a number of beautiful experiments in radiation optics, but these experiments usually give no information about the Abraham term as this term simply fluctuates out. So one is left with somewhat indirect verifications of this force, as in the radiation pressure of Jones {et al.} in the 1950's, testing the radiation pressure on a mirror immersed in a dielectric liquid. Now, there is a different  way to test the existence of ${\bf f}^{ \rm Aterm}$, namely to work with low (quasi-stationary) frequencies enabling one to observe the sinusoidal variation of the force directly. These kind of experiments were actually done by Walker {\it et al.} in 1he 1970's, using BaTiO$_3$ as a high-frequency dielectric (permittivity $\varepsilon \sim 3600$). The outcome of these experiments was convincing.  Now, in recent years there have appeared dielectric materials with giant permittivities, of order $10^5$ or even higher. It is therefore natural to consider the idea of Walker {\it et al.} anew, in order to test if this demanding experiment can be facilitated and give better accuracy. That is the main topic of the present paper. The positive outcome of these kinds of experiments  clearly supports the Abraham energy-momentum tensor at low frequencies. The  Minkowski tensor is unable to predict a torque at all.   The low-frequency and the high-frequency regimes are in this way highly contrasted, as it is obvious that in optical experiments the Minkowski tensor is by far the simplest and most convenient one to use.  We end this note by commenting upon the use of the Einstein-Laub tensor (1908) in explaining this experiment, and discuss also the influence from air friction.
\end{abstract}
 \maketitle

\section{Introduction}
\label{ChaptIntro}

The classic Abraham-Minkowski problem (for some original references see  \cite{abraham09,abraham10,minkowski10}), focuses on what is the correct, or at least the most convenient, expression for photon momentum propagating an a dielectric medium. To give a proper background for the present investigation, let us begin with the general expression for the basic electromagnetic  force density ${\bf f}$  in an isotropic dispersive fluid   with mass density $ \rho_m$, \cite{landau84,brevik79,brevik18},
\begin{align}
{\bf f}=&\rho{\bf E}+{\bf J\times B}  - \frac{1}{2}\varepsilon_0 E^2{\bf \nabla} \varepsilon
-\frac{1}{2}\mu_0H^2{\bf \nabla}\mu     +   \frac{1}{2}\varepsilon_0 {\bf \nabla}\left[ E^2\rho_m\frac{\partial \varepsilon}{\partial \rho_m}\right]+
\frac{1}{2}\mu_0 {\bf \nabla}\left[ H^2\rho_m\frac{\partial \mu}{\partial \rho_m}\right] \notag \\
& +\frac{n^2-1}{c^2}\frac{\partial}{\partial t}{\bf (E\times H)}. \label{1}
\end{align}
Here the constitutive relations are written in the form ${\bf D}= \varepsilon\varepsilon_0 {\bf E}, \,{\bf B}= \mu\mu_0 {\bf H}$, so that $\varepsilon$ and $\mu$ become nondimensional. The refractive index is $n=\sqrt{\varepsilon\mu}$. If the medium is nonpolar, the partial derivatives  $\partial \varepsilon/\partial \rho_m$  and    $\partial \mu/\partial \rho_m$     will be independent of whether they are  taken at constant entropy or at constant temperature. The Clausius-Mossotti relation is the applicable in the nonmagnetic case.

In the following we assume that the charge density $\rho$ and the current density $\bf J$ are zero, and that the medium is nonmagnetic. The  force expression above simplifies to
\begin{align}
{\bf f}=     -\frac{1}{2}\varepsilon_0E^2{\bf \nabla}n^2 +      \frac{1}{2}\varepsilon_0 {\bf \nabla}\left[ E^2\rho_m\frac{\partial n^2}{\partial \rho_m}\right]+
&  \frac{n^2-1}{c^2}\frac{\partial}{\partial t}{\bf (E\times H)}, \label{2a}
\end{align}
with $n^2=\varepsilon$.
The first term in the above expression is the one usually detected in radiation experiments in optics. As in such cases $n>1$, it means that radiation forces acting in inhomogeneous regions in the fluid act toward the optically thinner region. This kind of experiments, started by Ashskin and Dziedzic back in 1973 \cite{ashkin73}, has later been repeated under different circumstances, and are showing the effect very clearly. Some recent papers can be consulted in Refs.~\cite{casner03,wunenburger11,astrath14,capeloto16,zhang88,kundu17,zhang15,choi17}. This kind of force is evidently of importance, both from a fundamental and a practical viewpoint, especially in microfluidic contexts.

 The second term in Eq.~(\ref{2a})  is the electrostriction  term. This force gives rse to a local pressure increase in the fluid, but is usually not of practical importance since being spatial  gradient it can be transformed into a surface integral giving zero total force   if the body is surrounded by a vacuum.

  The classic experiment of Hakim and Higham \cite{hakim62} is here important. They measured the refractive index increase in a dielectric fluid when located in the region between two plates in the presence of a strong electric field. Theoretical considerations of this effect can be found at various places; cf. \cite{brevik79,lai89,zimmerli99,ellingsen11,partanen17a}. An interesting recent experimental confirmation of the dispersive effect is recently reported \cite{xi21}. These authors exploited the lack of axial symmetry in a single-mode optical fiber, and were able to measure the radiation force, including the electrostriction force, very accurately.

 The third member in Eq.~(\ref{2a}) is the famous Abraham term. It plays a central   in the discussions about the preferable energy-momentum tensor.  Experimental verifications of the Abraham term are scarce. There exist several claims in the literature, based upon experimental observations,  about "solving" the Abraham-Minkowski problem, but many of these are over-interpreting  the importance of the observations.  They can often be described using only the first term (the surface force) in Eq.~(\ref{2a}) (a closer discussion given in Ref.~\cite{brevik18}).

 It is now convenient to write down the Minkowski and Abraham energy-momentum tensors $S_{\mu\nu}^M$ and $S_{\mu\nu}^A$ (cf., for instance, Ref.~\cite{moller72}). The spatial part of Minkowski's tensor is
  \begin{equation}
 S_{ik}^{\rm M}= -E_iD_k-H_iB_k+\frac{1}{2}\delta_{ik}({\bf E\cdot D+H\cdot B}), \label{3}
 \end{equation}
  the energy flux density (Poynting vector) and momentum density are
 \begin{equation}
 {\bf S}^{\rm M}={\bf E\times H}, \quad {\bf g}^{\rm M}={\bf D\times B}, \label{4}
 \end{equation}
and the  energy density is
\begin{equation}
w^{\rm M}=\frac{1}{2}{\bf (E\cdot D+H\cdot B)}. \label{5}
\end{equation}

The Minkowski force density for a nonmagnetic medium thus becomes
\begin{equation}
{\bf f}^{\rm M}= -\frac{1}{2}\varepsilon_0 E^2{\bf \nabla} n^2. \label{6}
\end{equation}
In the Abraham case we have $S_{ik}^{\rm A}=S_{ik}^{\rm M},\, {\bf S}^{\rm A}={\bf S}^{\rm M}={\bf E\times H}, \, w^{\rm A}=w^{\rm M}$. The difference from the foregoing case occurs in the force density, which becomes
\begin{equation}
{\bf f}^{ A}= {\bf f}^{ M} +\frac{n^2-1}{c^2}\frac{\partial}{\partial t}{\bf (E\times H)}\mathrm{}. \label{7}
\end{equation}
The core of the Abraham-Minkowski problem thus lies in the Abraham term.   As mentioned, this  term is under optical circumstances small, and  rapidly fluctuating, and difficult to detect experimentally. As in the Minkowski case, the Abraham theory is unable to encompass the electrostriction forces.

The expression (\ref{1}) is thus the Abraham force density, with the   electrostriction and magnetostriction forces added. It is often called the Helmholtz force density.

\section{Updated version of the experiment of Walker et al.~\cite{walker75a,walker75b}}

We will henceforth be concerned with the Abraham term, called ${\bf f}^{\rm Aterm}$, under quasi-stationary  conditions, when the permittivity of the test medium is very large. It is then natural to revert to the symbol $\varepsilon$ instead of the optical symbol $n^2$, for the permittivity. Thus,
\begin{equation}
{\bf f}^{ A}= {\bf f}^{ M} +  {\bf f}^{\rm Aterm}, \label{8}
\end{equation}
with
\begin{equation}
{\bf f}^{\rm Aterm} = \frac{\varepsilon-1}{c^2}\frac{\partial}{\partial t}{\bf (E\times H)}. \label{9}
\end{equation}
The experiment of Walker {\it et al.} \cite{walker75a,walker75b} measured the torque on a torsional pendulum (annular cylinder) suspended in the gravitational field. The frequencies were low, $\omega/2\pi \sim 0.4~$Hz, in resonance with the applied Abraham term (\ref{9}) in order to maximize the torque and the angular deflection here called $\phi$.  A  strong vertical constant magnetic field, $B_0 = \mu_0H_0 \sim 1~$T, was created by an electromagnet, and a harmonically varying radial electric field
\begin{equation}
E= \frac{V_0}{r\ln (b/a)} \sin \omega t\label{10}
\end{equation}
 was imposed between the aluminium-coated inner and outer cylindrical surfaces $r=a$ and $r=b$. Here $V_0 \sim 260~$V  was the voltage across the disk. The material in the  disk (height $L$) was BaTiO$_3$, with permittivity $\varepsilon =3620$. The suspended system was heavy, in total about 800 kg including the magnet. The pendulum was suspended by a thin tungsten fiber. The experiment was performed at a low pressure,   $10^{-5}~$torr.

 We can now from Eqs.~(\ref{9}) and (\ref{10})  write the axial torque $T$ in the form
 \begin{equation}
 T = K\omega \cos \omega t, \label{11}
 \end{equation}
 where
 \begin{equation}
 K= \frac{\pi \varepsilon_0(\varepsilon-1)L}{\ln(b/a)}(b^2-a^2)B_0V_0. \label{11a}
 \end{equation}
 When applied to the pendulum, $T$ can be written as
 \begin{equation}
 T= I(\ddot{\phi}+2\lambda \dot{\phi}+\Omega^2 \phi), \label{13}
 \end{equation}
 where $I$ is the moment of inertia, $I= (1/2)\rho_m L(b^4-a^4)$, $\lambda$ is the damping coefficient, and $\Omega$ is the eigenfrequency in the absence of any damping. (For comparison purposes, note that the symbols $\delta$ and $\gamma$ in Ref.~\cite{walker75a} correspond to $\delta= 2I\lambda $ and $\gamma = I\Omega^2$.)

 We  will seek only the stationary solution of  the equation of motion
 \begin{equation}
 \ddot{\phi}+2\lambda \dot{\phi}+\Omega^2 \phi= \frac{K}{I} \omega  \cos \omega t.  \label{14}
 \end{equation}
Assuming  the form
\begin{equation}
\phi= b\cos (\omega t+\delta), \label{15}
\end{equation}
with $b$ and $\delta$ as constants, we derive
\begin{equation}
b= \frac{(K/I)\omega}{\sqrt{(\Omega^2-\omega^2)^2+4\lambda^2\omega^2}}, \quad \tan \delta= \frac{2\lambda \omega}{\omega^2-\Omega^2}. \label{16}
\end{equation}

 In general, the  deflection $\phi$ lags behind the applied torque $T$. The angle $\delta$ is negative. Far from resonance on the side $\omega < \Omega$, one gets $\delta \rightarrow 0$, whereas on the other side $\omega >\Omega$, $\delta \rightarrow -\pi$. The passage of $\delta$ through $-\frac{1}{2}\pi $ occurs when $\omega = \Omega$.

 Of main interest is the region near resonance. We put $\omega = \Omega +\xi$ with $|\xi| \ll 1$, and assume  that $\lambda \ll \Omega$. Then approximately
 \begin{equation}
 b= \frac{K/2I}{\sqrt{\xi^2+\lambda^2}}, \quad \tan \delta = \frac{\lambda}{\xi}, \label{16a}
 \end{equation}
 showing that the amplitude $b$ does not now depend on the driving frequency $\omega$.
 In the actual experiment, two values of the driving frequency were investigated, $\omega/2\pi = 0.24$ and $0.48~$Hz. In the present  analysis, we shall simplify to the case $\xi=0$, $\delta = -\frac{1}{2}\pi$, whereby from the formulas above we get the following  expression for the deflection,
 \begin{equation}
 \phi= \frac{K}{2I\lambda}\sin \omega t. \label{17}
 \end{equation}
It is convenient to insert the same geometric data as in  the experiment, $b=2.63~$cm, $a=0.40~$cm, $L=1.99~$cm, cylinder  mass $M=217~$g, $I= 779~$ gcm$^2$, time constant $\tau=1/\lambda =1~$hour, $B_0= 1~$T, $V_0 = 260~$V, keeping $\varepsilon$ and $\omega$ unspecified. That leads to
\begin{equation}
K= 5.16(\varepsilon-1)\times 10^{-14}~\rm{N~m~ s}, \label{18}
\end{equation}
and so
\begin{equation}
\phi = 1.2  (\varepsilon-1)\times 10^{-6}\sin\omega t. \label{19}
\end{equation}
Thus with $\varepsilon = 3620$, the permittivity used in the experiment, we get a maximum deflection of $\phi_0 = 4.3\times 10^{-3}~$rad. The maximum torque is given by Eq.~(\ref{11}) as $T_0= K\omega$. Assuming $\omega/2\pi= 0.4~$Hz,  this yields $T_0 = 4.7 \times 10^{-10}~$Nm. These are theoretical formulas. The experimenters, carefully taking into account corrections, were able to get agreement with the theoretical prediction to within $\pm 10\%$.

This is  an important classic experiment, showing unambiguously the existence of the Abraham force. Note that the prediction from the Minkowski force (\ref{6}) is quite different, as it corresponds to a {\it radial} force, with no component in the azimuthal direction at all, thus no torque.

We have now come to a main point of this paper, namely to propose a repetition of the experiment of Walker {\it et al.} taking advantage of  the impressive improvements in the fabrication of high-permittivity materials in recent years. They are often called colossal permittivity materials (CP), where  the permittivities can exceed $10^3$ by far. They are of obvious interest for modern electronics and energy storage systems. A useful reference is the review article of Wang {\it et al.} \cite{wang19}. The most well-known nonferroelectric CP material is CaCu$_3$Ti$_4$O$_{12}$  (called CCTO), being stable under a wide temperature range including room temperatures, and being practically constant at the frequencies of interest here. (There are even other types of materials called supercapacitors, containing liquid electrolytes, where the permittivities can be as large as about $10^9$ \cite{cortes15}.)

For definiteness, assume a CCTO single crystal  with $\varepsilon = 10^5$  (Fig.~11 in Ref.~\cite{wang19}) used in the Walker {\it et al.} experiment, the other parameters being the same as above. It then follows that
\begin{equation}
\phi= 0.12 \sin \omega t, \label{20}
\end{equation}
and with $\omega/2\pi = 0.4~$Hz,
\begin{equation}
T_0= 1.2\times 10^{-8}~\rm{Nm}. \label{21}
\end{equation}
This indicates an essential simplification of the classic experiment: the angular deflection, as well as the torque, become amplified by a factor of about 27. In our opinion this significant change makes a repetition of the experiment highly worth considering.

\section{Final remarks}

Whereas the conclusion of the present  investigation is obvious from the previous section, we will  add three remarks.

\bigskip

\noindent 1. {\it Comparison with the Enstein-Laub prediction}.  In addition to the best known Abraham and Minkowski alternatives for the electromagnetic energy-momentum tensor in media, there exist also other alternatives, like the one proposed by Peierls, for instance \cite{peierls76}. A more well-known alternative is the proposal of Einstein and Laub back in 1908 \cite{einstein08}. Calling  this energy-momentum tensor $S_{\mu\nu}^{\rm E}$, we have for its spatial part
\begin{equation}
S_{ik}^{\rm E}= -E_iD_k-H_iB_k +\frac{1}{2}\delta_{ik}(\varepsilon_0 E^2+\mu_0H^2), \label{22}
\end{equation}
and its energy flux density and momentum density are
\begin{equation}
{\bf S}^{\rm E}= {\bf E\times H}, \quad {\bf g}^{\rm E}= \frac {1}{c^2}{\bf E\times H}. \label{24}
\end{equation}
We  define the corresponding energy density as
\begin{equation}
w^{\rm E}= \frac{1}{2}(\varepsilon_0E^2+\mu_0 H^2), \label{25}
\end{equation}
although this quantity was not explicitly given by Einstein and Laub. (In  this way we make the energy-momentum tensor identical to one of the alternatives derived by de Groot and Suttorp \cite{deGroot72}.)

The general expression for the energy density ${\bf f}^{E}$ is, when $\rho = {\bf J}=0$ (cf. also Sec. 1.2 in \cite{brevik79}),
\begin{equation}
{\bf f}^{E} ={\bf (P\cdot \nabla)E} + \mu_0{\bf (M\cdot \nabla)H} +\mu_0{\bf \dot{P}\times H}+\frac{1}{c^2}{\bf E\times \dot{M}}, \label{26}
\end{equation}
where ${\bf D}= \varepsilon_0{\bf E}+{\bf P}$.
If the medium is isotropic and nonconducting, the following rewriting is useful,
\begin{equation}
{\bf f}^{E} = {\bf f}^{A}+ \frac{1}{2}{\bf \nabla (E\cdot P}
 + \mu_0{\bf H\cdot M)}.\label{27}
\end{equation}
In the present case where ${\bf M}=0$, the expression (\ref{26}) is reduced to two terms. The first term containing ${\bf E}$ has no azimuthal component, and so we are left with the single effective term
\begin{equation}
{\bf f}^{E}   = \mu_0{\bf \dot{P}\times H} = \frac{\varepsilon-1}{c^2}\frac{\partial}{\partial t}{\bf (E\times H)}, \label{28}
\end{equation}
since $\bf H$ is assumed constant. Thus ${\bf f}^{\rm E} = {\bf f}^{\rm Aterm}$, where the last quantity given by Eq.~(\ref{9}). The Abraham and the Einstein-Laub expressions predict the same result in this case. Also, in the recent experiment of angular symmetry of the radiation force in a solid dielectric \cite{xi21}, the investigators give preference to the Einstein-Laub formulation.

Does this imply that the Einstein-Laub formulation is fully equivalent to the Abraham formulation? The answer is no. This is seen if we go back to  the electrostrictive generalization of the Abraham theory, that means, to the Helmholtz force ${\bf f}^{\rm H}$, which follows from Eq.~(\ref{1}) as
\begin{equation}
{\bf f}^{\rm H} =  - \frac{1}{2}\varepsilon_0 E^2{\bf \nabla} \varepsilon
   +   \frac{1}{2}\varepsilon_0 {\bf \nabla}\left[ E^2\rho_m\frac{\partial \varepsilon}{\partial \rho_m}\right]. \label{29}
\end{equation}
As already mentioned, in  the  Hakim-Higham experiment \cite{hakim62} the electrostrictive excess pressure $\Delta p$ was measured in a dielectric liquid in the strong field region between two plates completely immersed in the liquid. From Eq.~(\ref{29}) it follows that the Helmholtz pressure $\Delta p^{\rm H}$ in the interior homogeneous region where ${\bf \nabla}\varepsilon=0$ is
\begin{equation}
\Delta p^{\rm H} = \frac{1}{6}\varepsilon_0 E^2(\varepsilon-1)(\varepsilon +2), \label{30}
\end{equation}
where use has been made of the Clausius-Mossotti relation.
The corresponding Einstein-Laub excess pressure $\Delta p^{\rm E}$ follows directly from Eq.~(\ref{27}) as
\begin{equation}
\Delta p^{\rm E} = \frac{1}{2}\varepsilon_0 E^2(\varepsilon-1). \label{31}
\end{equation}
The experiment showed agreement with the prediction (\ref{30}), to $\pm 5$ per cent,  and disagreed  with (\ref{31}). The Abraham alternative, when in an electrostrictive situation extended to the Helmholtz expression, thus stands out as the preferable option on the fundamental level.

\bigskip

\noindent 2. {\it Influence from air friction}. In air surroundings, there will be viscous damping forces from the air viscosity. It is of interest to estimate the magnitude of this correction. We go back to the Abraham theory, and assume for simplicity the  viscous shear stress on the surface of the sample can be evaluated as if the cylinder were long ($\L \rightarrow \infty$). Moreover, we assume in connection with friction the cylinder to be uniformly rotating with constant angular velocity $\dot{\phi}$.

From the Navier-Stokes equation we find  the angular velocity $v$ of air outside and inside the annular test body to be
\begin{equation}
v= \frac{\dot{\phi} b^2}{r}, \quad r \geq b, \label{32}
\end{equation}
\begin{equation}
v= \dot{\phi} r, \quad r \leq a. \label{33}
\end{equation}
The inner air volume $0<r<a$ thus rotates as a solid body, and there is no shear stress on the surface $r=a$. The shear stress $\sigma_{r\phi}$ on the surface $r=b$ is
\begin{equation}
\sigma_{r\phi} =\eta \left( \frac{\partial v}{\partial r}-\frac{v}{r}\right)_{r=b}= -2\eta \dot{\phi}, \label{34}
\end{equation}
where $\eta$ is the shear viscosity for air. The viscous torque on a length $L$ of the cylinder thus becomes
\begin{equation}
T_{\rm viscous}= -4\pi \eta \dot{\phi} b^2L. \label{35}
\end{equation}
The expression for (\ref{13}) for the torque now gets modified, and can be written as
\begin{equation}
T= I[\ddot{\phi}+2(\lambda + \lambda_{\rm viscous}) \dot{\phi}+\Omega^2 \phi], \label{36}
\end{equation}
with
\begin{equation}
\lambda_{\rm viscous}=\left( \frac{2\pi \eta}{I}\right) b^2L. \label{37}
\end{equation}
The equation of motion becomes
\begin{equation}
 \ddot{\phi}+2(\lambda + \lambda_{\rm viscous}) \dot{\phi}+\Omega^2 \phi= \frac{K}{I} \omega  \cos \omega t, \label{37a}
 \end{equation}
 and the solution is given by Eqs.~(\ref{16}) and (\ref{16a}) with $\lambda$ replaced by $(\lambda +\lambda_{\rm viscous})$.

Assume first atmospheric pressure, for which $\eta =1.8\times 10^{-5}~$Pa\,s $= 1.8\times 10^{-4}~$dyn\,s\,cm$^{-2}$.  Taking as before $I= 779~$g\,cm$^2$,\, $L=1.99~$cm,\,$b=2.63~$cm, we obtain
\begin{equation}
\lambda_{\rm viscous} = 2.0\times 10^{-5}~\rm{s}^{-1}. \label{38}
\end{equation}
As mentioned above, the authors estimated the time constant to be $
\tau=1/\lambda= 1~$hour. With these values,
\begin{equation}
\frac{\lambda_{\rm viscous}}{\lambda} =0.072. \label{39}
\end{equation}
This is a quite small correction. As pointed out in their paper, the time constant might in reality have been higher, perhaps around 2 hours. That would in case double the ratio (\ref{39}). As is shown by Eq.~(\ref{37}), the running of the experiment at the low pressure of  $10^{-5}~$torr, as was actually done, was beneficial as a lower air density  implied a lower viscosity $\eta$, thus a lower $\lambda_{\rm viscous}$.

\bigskip
\noindent 3. {\it An optical analog: use of whispering gallery modes.} As mentioned, at optical frequencies the direct observation of Abraham's fluctuating force is practical impossible; the fluctuations become averaged out. As our last remark, we wish however to emphasize that a critical experimental of this force can be achievable after all, by an {\it intensity modulation} of a laser beam. This topic was actually considered in some detail earlier, in Ref.~\cite{brevik10}, but we will comment briefly on it also here. The subject involves ordinary dielectric media of refractive index $n$ only; thus has no relationship to giant permittivities. The reason why we nevertheless consider it here, is for comparison purposes.

In practice, the optical variant of our problem involves dielectric microresonators, primarily of spherical form. Let us consider such a compact sphere with radius $a \sim 50-100~\mu$m. When fed by an external source , whispering gallery modes can be  built up in the interior with large azimuthically    circulating power, of order $P \sim 100~$W or more, concentrated near the rim. If the sphere is suspended in the gravitational field, it can execute torsional oscillations, essentially in the same way as above. Assume $P=P_0\cos\omega_0t$, where $\omega_0$ is the modulation frequency. The circulating energy flux density varies in the same way,
\begin{equation}
S= S_0\cos \omega_0t, \label{41}
\end{equation}
where the complicated expression for $S_0$ as a function of the input parameters is given in Ref.~\cite{brevik10}. The Abraham term (\ref{9}) predicts an azimuthal force density component
\begin{equation}
f = -\frac{n^2-1}{c^2}\omega_0S_0\sin \omega_0t, \label{42}
\end{equation}
and a corresponding axial torque on the sphere,
\begin{equation}
T= \int rf\sin\theta dV, \label{43}
\end{equation}
with $dV= r^2\sin\theta dr d\theta d\phi$.

We now omit corrections from air friction, and put for definiteness $a=100~\mu$m, $P_0= 100~$W. The moment of inertia about the $z$ axis is $I= (2/5)Ma^2$, where $M =4~\mu$g is the mass. From earlier experimental tests   of the equivalence principle \cite{hou03,schlamminger08}, we adopt the relationship $I\Omega^2 \sim 10^{-9}~$N\,m/rad, where $\Omega$ as before is the undamped angular frequency of the test body. This leads to
\begin{equation}
\Omega \sim 10^3 \, \rm{rad\, s}^{-1}. \label{44}
\end{equation}
The torque varies with time as $T=T_0\sin \omega_0t$, and with the given data we estimate for the amplitude
\begin{equation}
T_0 \sim 10^{-23}\omega_0~\rm{N\,m\,s}. \label{45}
\end{equation}
This is very small. Realistic values for the modulation frequency are moderate;  even with the relatively high value of $\omega_0 = 10^6~$rad/s we would obtain only
\begin{equation}
T_0 \sim 10^{-17}~\rm{N\,m}. \label{46}
\end{equation}
This is one order of magnitude smaller than the value $10^{-16}~$N\,m obtained by Beth \cite{beth36} in his classic experiment on photon angular momentum.

The corresponding values of the angular deflection are also small; we estimate for the amplitude
\begin{equation}
\phi_0 \sim 10^{-8}~\rm{rad}. \label{47}
\end{equation}
In conclusion, we find a comparison between Eqs.~(\ref{20}), (\ref{21}) in the macroscopic case, and Eqs.~(\ref{46}), (\ref{47}) in the microscopic case, to be instructive. The two cases aim at measuring the same Abraham force, but under widely different conditions.

\end{document}